\documentclass[final]{anthology-ch} 

\usepackage{booktabs}
\usepackage{graphicx}

\title{Echoes of Ideology – Toward an Audio Analysis Pipeline to Unveil Character Traits in Historical Nazi Propaganda Films}

\author[1]{Nicolas Ruth}[
  orcid=0000-0002-1126-6280
]

\author[1]{Manuel Burghardt}[
  orcid=0000-0003-1354-9089
]

\affiliation{1}{Computational Humanities Group, Leipzig University, Leipzig, Germany}


\pubyear{2025}
\conferencename{Proceedings of ADHO Digital Humanities conference (DH2025), 2025, Lisbon, Portugal}

\begin{document}

\maketitle

\begin{abstract}
This study investigates the use of computational audio analysis to examine ideological narratives in Nazi propaganda films. Employing a three-step pipeline—speaker diarization, audio transcription, psycholinguistic analysis—it reveals ideological patterns in characters. Despite current issues with speaker diarization, the methodology provides insights into character traits and propaganda narratives, suggesting scalable applications.
\end{abstract}

\section{Introduction}

The Nazi regime's exploitation of film as a propaganda tool underscores its belief in cinema's unparalleled power to shape minds and spread ideology. Implementing Joseph Goebbels' concept of a "revolution of the mind" (Leiser, as cited in \cite{hardinghaus2008}), the Nazis subjected all art and culture to their worldview, establishing a centralized and tightly controlled film industry to produce ideologically driven cinema, even advocating for a new film style \cite{giesen2005hitlerjunge}. Today, a total of 44 films remain officially classified as Vorbehaltsfilme (restricted films) because of their particularly strong ideological and historical significance. Preserved by the Friedrich-Wilhelm-Murnau Foundation, these films are accessible only under strict academic research agreements. Interestingly, many other films from the same period, like the German classic "Die Feuerzangenbowle" (English: The Fire-Tongs Bowl) which still airs every Christmas, remain widely available. There is no clear definition that can be used to decide which Nazi films should be classified as unproblematic and which should not. On the contrary, the entire history of film censorship in post-war Germany has been heavily shaped by societal changes and varying influences, such as those stemming from different occupation zones \cite{hoppe2021diskurs}. There is still an ongoing debate over the distinction between propaganda and non-propaganda films. The field of history has also recently renewed its engagement with the legacy of Nazi-era cinema. In early 2025, one of Germany’s leading historical research institutes launched a new project entitled "Screens", which will examine various elements of these films and their ideological foundations.\footnote{https://www.ifz-muenchen.de/aktuelles/artikel/screens-spielfilme-als-zeithistorische-quelle} In this study, we aim to analyze a selection of the restricted Vorbehaltsfilme to understand how they construct their ideological narratives. These include Jud Süß (1940) (English: Süss, the Jew), Hitlerjunge Quex (1933) (English: Hitler Youth Quex) and Kopf hoch, Johannes! (1941) (English: Chin up, Johannes!). By employing a computational analysis approach, we aim to uncover ideological patterns within these films and eventually extend this method to the hundreds of publicly available Nazi-era films to reveal how much "hidden" ideology they may actually contain.

\section{Methodological Considerations} 
In film studies, it is widely recognized that film is a multimodal \cite{bateman2013multimodal} and highly complex phenomenon that embodies a visual as well as an audio layer, with the latter further divided into spoken language, music, and sound effects. We argue that the visual layer has been widely studied in digital film studies, leading to concepts such as “distant viewing” \cite{arnold2019distant} and “deep watching” \cite{bermeitinger2019deep}, which use modern deep learning technologies and contribute to what has been called the "digital visual turn" \cite{wevers2019turn}. However, we believe that a significant methodological gap remains in the computational analysis of audio, especially in historical films. This is why we propose and evaluate a three-step analytical workflow specifically for the audio analysis of historical Nazi films. For now, our focus is on character dialogs, as we believe ideological patterns are primarily conveyed through film characters and their spoken interactions.
The analysis of the conceptualization of film characters in Nazi-era cinema presents a range of research opportunities. Two key areas include: the analysis of cultural promotion of ethnically defined in- and out-groups, where (e.g. antisemitic) portrayals assigned discriminatory traits \cite{friesel2017sprache}, and the depiction of femininity, which was viewed as a reflection of cultural identity in Nazi-Germany \cite{frietsch2015mediale}. Insights may be gained by automatically comparing character portrayals to manually selected references, and by applying voice gender detection within the audio analysis pipeline.
Our computational analysis of character traits is methodologically related to the field of “psycholinguistics”. \cite{herderich2024psycho} describe the emerging use of computer-based NLP methods to quantitatively assess psychologically-grounded character traits, bridging the gap between qualitative and quantitative research. These methods include techniques like topic modeling, which are applied to traditional psychological interviews, or dictionary-based approaches such as LIWC \cite{boyd2022liwc22} and Empath \cite{fast2016empath}, which infer traits from linguistic elements. Some researchers even argue that language may be a stronger indicator of personality than traditional psychological self-reports \cite{boyd2017personality, boyd2020personality}. For the computational analysis of large language corpora, recent advances in transformer-based language models show that these consistently outperform traditional dictionary-based approaches, for example for tasks such as sentiment analysis \cite{borst2023zero}. Building on this work, we employ an LLM-based approach for analysis of character traits of the film characters. Although fictional characters use scripted language, this is precisely of interest here, as we are concerned with how character portrayal was shaped by National Socialism.

\section{Audio Analysis Workflow}
\begin{figure}[h!]
  \centering
  \includegraphics[width=\linewidth]{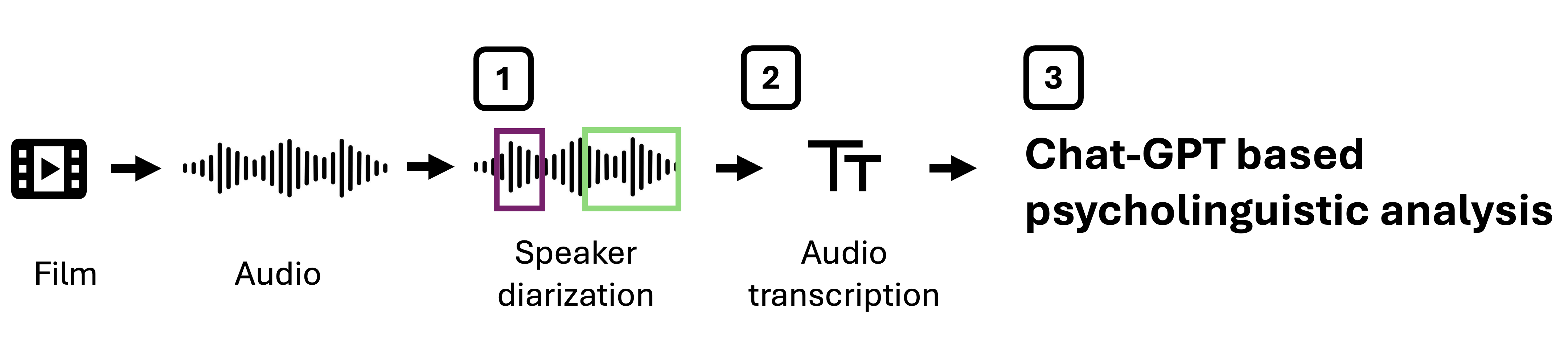}
  \caption{Overview of the three-step analysis workflow.}
  \label{fig:overview}
\end{figure}

We suggest a three-step process (see Fig. \ref{fig:overview}) for automatically transforming spoken film dialog into a character analysis and evaluate state-of-the-art deep learning models for each step.

First, the audio layer is extracted from the film and passed to a step dedicated to (1) speaker diarization, which means the identification of different speakers. This process involves voice activity detection (VAD), computing speaker embeddings, and clustering speech segments by speaker identity. This task is particularly challenging, as we are dealing with historical film audio from 1933 to 1945. For our analysis, we evaluated two popular open-source tools: Pyannote's\footnote{https://github.com/pyannote/pyannote-audio} speaker diarization and Nvidia Nemo\footnote{https://docs.nvidia.com/nemo-framework/user-guide/latest/nemotoolkit/asr/speaker\_diarization/intro.html} \cite{fischbach2024comparative}. The next step in our pipeline is (2) audio transcription, which uses automatic speech recognition (ASR) to convert the audio into text associated with the identified speaker. We evaluated OpenAI’s Whisper \cite{radford2023whisper} in four versions: Whisper-medium, Whisper large-v2, and German fine-tuned versions\footnote{https://huggingface.co/bofenghuang/whisper-medium-cv11-german \& https://huggingface.co/bofenghuang/whisper-large-v2-cv11-german} by Bofeng Huang. Whisper is considered state-of-the-art, noted for its robustness to background noise \cite{kuhn2024asr}. Finally, the collected speeches for each speaker undergoes a (3) “psycholinguistic” analysis of character traits using a GPT-based approach. This process aggregates the speech segments for each (main-) speaker and submits them as a list to gpt- 3.5-turbo-0125 via the API with the following prompt (translated to English from the original German prompt):

\begin{figure}[h!]
  \centering
  \includegraphics[width=\linewidth]{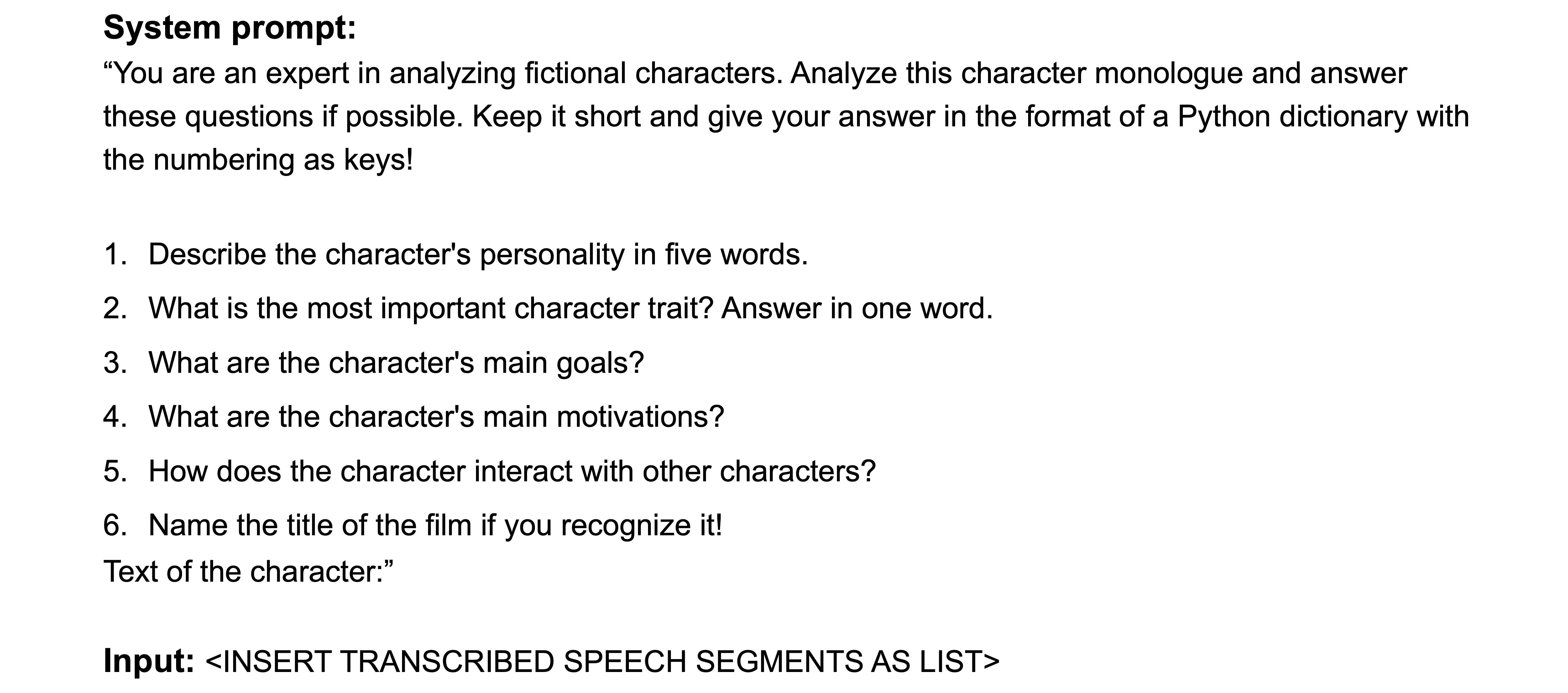}
  \caption{System prompt and input of the GPT-based character analysis.}
  \label{fig:prompt}
\end{figure}

The smaller-model (gpt-3.5-turbo-0125) was chosen because of the cost considerations of the API. Main speakers are selected by creating a speech frequency histogram (for speeches >2s) and choosing those with frequencies above the upper fence, ensuring adaptability to the film’s speech patterns and sufficient material for psycholinguistic analysis.

\section{Evaluation of methods}

\subsection{Step 1: Speaker diarization}

While typically speaker recognition systems are designed for rather controlled settings, this study faces added complexities: historical German audio, movie audio with background noise, uncertain speaker numbers, and overly long audio signals. Additionally, the complex framework's models and their individually configurable steps introduce further methodological difficulties and potential errors. The audio-based speaker recognition system was tested through various configurations of Pyannote, including its standard version, two voice isolation techniques\footnote{https://pypi.org/project/noisereduce/ \& https://huggingface.co/speechbrain/sepformer-dns4-16k-enhancement}, a German segmentation model\footnote{https://huggingface.co/diarizers-community/speaker-segmentation-fine-tuned-callhome-deu}, a segmentation model fine-tuned on “Jud Süß” and a custom method correcting speaker confusion via face embedding clusters. Additionally, the Nemo (general config, static 30 speakers) model was evaluated and outperformed all Pyannote configurations with an average diarization error rate (DER) of 58.84. DER measures speaker diarization performance as the percentage of incorrectly attributed speech time, including false detections, missed speech, or incorrect clustering, and can exceed 100\% due to overlapping speech \cite{fiscus2006DER}. This step towards full automation presents the biggest challenge to be solved in the future, which is why the subsequent steps in our workflow rely on manually created ground truth for speaker recognition. For this task, there are various approaches for complex video material, including cross-modal models that need to be tested with the data \cite{sharma2023cross, cheng2025mulitmodalSD}.

\subsection{Step 2: Transcription}

Regarding the automatic speech recognition the models show a high robustness for background noises and a strong capability to capture the correct punctuation marks, therefore providing insight into the meaning of a sentence. The main failures are caused by older German vocabulary (for instance, Durchlaucht, Staatsstreich, Rabbuni), proper names, dialects/languages (for instance, Yiddish, Berlin dialect) and very quiet speech that provokes ASR-hallucinations. The German fine-tuned models perform slightly better with specific German words like Jude, Jid, and Jiden. However, both large variants are generally effective and can be used effectively. 
The following evaluation of the GPT-based analysis indicates that the transcription results – despite minor errors – are sufficient for the intended procedure. Studies, such as \cite{wang2024resilience}, have shown that ASR-errors and typographical errors can decrease the performance of LLMs. However, our application appears to be resilient to such issues as the ASR-error rate is low enough.

\subsection{Step 3: ChatGPT-Analysis}

The first insight into the GPT-based analysis is that the model does not recognize the films based on the provided character speech, as tested in question 6 of the prompt (see Fig. \ref{fig:prompt}). The only exceptions are when a rather famous main character is directly named in the speech, which is not necessarily the named character themself. This happens in the case of Heini Völker from “Hitlerjunge Quex” or Joseph Süß Oppenheimer in “Jud Süß”. Apart from these cases the model hallucinates random movie names indicating the analysis results are not based on prior knowledge about the films.

\begin{figure}[h!]
  \centering
  \includegraphics[width=\linewidth]{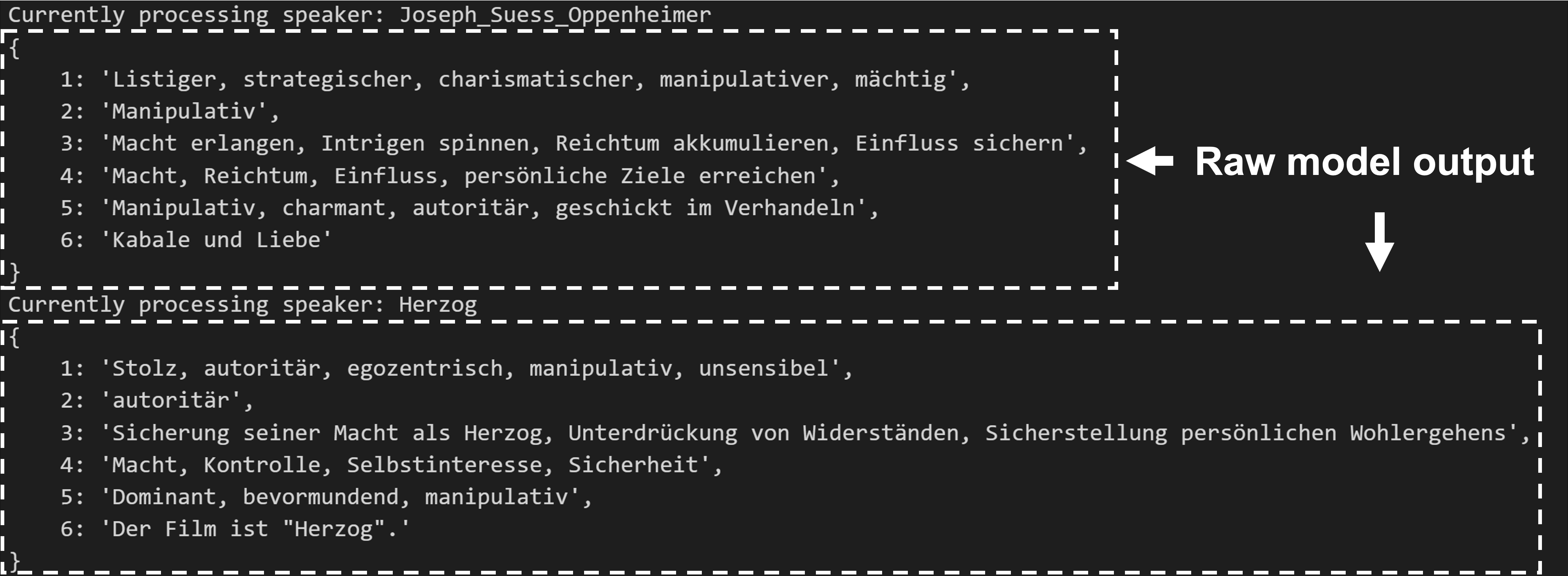}
  \caption{Output of GPT-based character analysis for the characters “Joseph Süß Oppenheimer” and the Duke, later translated to english for the paper.}
  \label{fig:answer}
\end{figure}

\begin{figure}[h!]
  \centering
  \includegraphics[width=\linewidth]{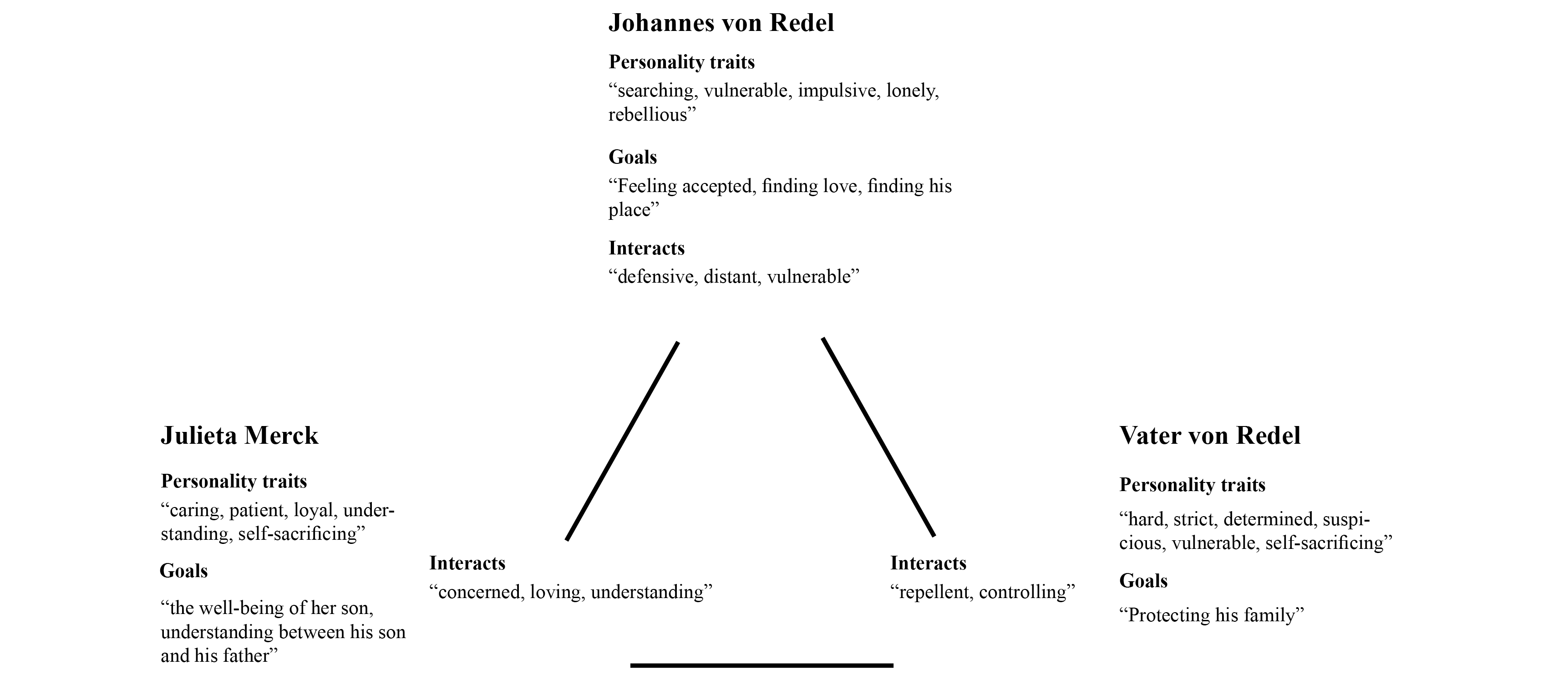}
  \caption{Character triangle of Julieta Merck, Johannes von Redel, and Johannes's father in the Nazi youth film “Kopf hoch, Johannes!”.}
  \label{fig:dreieck}
\end{figure}

Figure \ref{fig:answer} illustrates the direct analysis results for two characters from the anti-Semitic propaganda film Jud Süß. The approach proves to be accurate for analyzing characters, providing deep insights into their personalities based solely on speech passages. The analysis of Joseph Süß Oppenheimer describes him as a cunning, strategic, charismatic, and powerful character who manipulates, charms, and authoritatively interacts with others to gain power, wealth, and influence. These analysis results align with the scholarly literature on the film, which identifies Süß Oppenheimer's portrayal as a caricatured antisemitic personality \cite{niven2022jud}. The depiction of the Duke as proud, authoritarian, and egocentric, acting dominantly, patronizingly, and manipulatively to secure his power and personal welfare, can be considered accurate as well.
An impressive example from a different film is the character triangle of Julieta Merck, Johannes von Redel, and Johannes's father in the Nazi youth film “Kopf hoch, Johannes!”. The literature describes the major character dynamic as follows: Johannes von Redel, who returns to Germany after his mother's death, struggles to adjust and is distant from his rough, emotionally detached father. His aunt, Julieta Merck, attempts to bridge the rift between them \cite{giesen2005hitlerjunge}. The character triangle from the literature has been manually visualized in Figure \ref{fig:dreieck}, enhanced with the translated results on personality traits, goals, and interactions from the automated analysis, demonstrating the approach's ability to capture complex character traits.

\section{Conclusion}

It shows that the proposed workflow for the audio analysis of historical German film material requires further development in the area of speaker diarization, but already works very well for the tasks of audio transcription and character analysis. Future perspectives for speaker recognition include, in addition to the multimodal approaches already mentioned, a more in-depth focus on the audio data, involving specific fine-tuning and targeted model development based on historical audio samples. Current advancements in audio-based deep learning models present promising opportunities, especially if the strengths of models like Whisper, as a dominant ASR model, could be applied to other audio tasks. The audio transcription works robustly and sufficiently accurate. The identification of key character traits in the films, which in our tests was largely consistent with existing analyses in the respective literature, provides an opportunity for a quantitative analysis of character development in Nazi-era films, offering insights into Nazi ideology that can be scaled up to a larger corpus of films from that era, where we assume more hidden ideologies. For large-scale analyses of character traits, a closed vocabulary for GPT or word embedding systems will be pursued. Ultimately, we aim for a more holistic view on ideologies in Nazi films and plan to also adopt SOTA methods for a distant viewing of the visual layer, as is showcased in current tools such as Vian \cite{halter2019vian}, Tibava \cite{springstein2023tib} or Zoetrope \cite{liebl2023zoetrope}.

\section*{Ethics Statement}

This study involves the analysis of Vorbehaltsfilme (restricted Nazi propaganda films), which contain antisemitic, racist, and otherwise harmful content. These films are preserved by the Friedrich Wilhelm Murnau foundation and made available solely for academic research under controlled conditions. All analyses in this paper were conducted within a critical scholarly framework aimed at understanding the mechanisms of ideological construction in historical cinema. No material derived from the films is publicly distributed, and no attempt has been made to reproduce or disseminate harmful content outside of this research context. The study adheres to ethical principles of responsible data handling and historical sensitivity, acknowledging the continued impact of these materials and their potential for misuse.

\section*{Acknowledgements}
We thank the Friedrich Wilhelm Murnau foundation for granting access to the digitized Vorbehaltsfilme used in this research. Their support was essential in enabling the development and evaluation of the proposed audio analysis pipeline.

\printbibliography

\appendix

\end{document}